\documentclass[10pt]{IEEEtran}

\usepackage{cite}
\usepackage{graphicx}
\usepackage{amsmath}
\usepackage{array}
\usepackage[caption=false]{subfig}
\usepackage{amssymb}
\usepackage{xcolor}
\usepackage{url} 
\usepackage{multirow}
\usepackage{multicol}
\usepackage{comment}
\usepackage{tabu}
\usepackage{float}

\begin{document}
%\title{ DeepMI: Deep Multi-lead ECG Fusion for Occurrence-time Prediction of Myocardial Infarction}
\title{ DeepMI: Deep Multi-lead ECG Fusion for Identifying Myocardial Infarction and its Occurrence-time}

\author{Girmaw Abebe Tadesse,
Hamza Javed, 
Yong Liu, 
Jin Liu, 
Jiyan Chen,\\
Komminist Weldemariam, and 
Tingting Zhu
        
%         \footnote{$^1$IBM Research - Africa  \\
% $^2$University of Oxford, UK \\
% $^3$Guangdong Cardiovascular Institute,  China\\
% $^*$ Corresponding Author: \textcolor{blue}{girmaw.abebe.tadesse@ibm.com} }

% <-this % stops a space
\IEEEcompsocitemizethanks{
\IEEEcompsocthanksitem G. A. Tadesse and K. Weldemariam are with IBM Research - Africa. H. Javed and T. Zhu are with the University of Oxford, UK. Y. Liu, J. Liu and J. Chen are with Guangdong Cardiovascular Institute,  China.\protect
% E-mail: see http://www.michaelshell.org/contact.html
\IEEEcompsocthanksitem  Corresponding author: \textcolor{blue}{girmaw.abebe.tadesse@ibm.com}
}% <-this % stops a space

% \thanks{Manuscript received April 19, 2005; revised August 26, 2015.}
}

\IEEEtitleabstractindextext{%
\begin{abstract}

 Myocardial Infarction (MI) has the highest mortality of all cardiovascular diseases (CVDs). Detection of MI and information regarding its occurrence-time in particular, would enable timely interventions that may improve patient outcomes, thereby reducing the global rise in CVD deaths. Electrocardiogram (ECG) recordings are currently used to screen MI patients. However, manual inspection of ECGs is time-consuming and prone to subjective bias. Machine learning methods have been adopted for automated ECG diagnosis, but most approaches require extraction of ECG beats or consider leads independently of one another.  We propose an end-to-end deep learning approach, DeepMI, to classify MI from normal cases as well as identifying the time-occurrence of MI (defined as acute, recent and old), using a collection of fusion strategies on 12 ECG leads at data-, feature-, and decision-level. In order to minimise computational overhead, we employ transfer learning using existing computer vision networks. Moreover, we use recurrent neural networks to encode the longitudinal information inherent in ECGs. We validated DeepMI on a dataset collected from 17,381 patients, in which over 323,000 samples were extracted per ECG lead. We were able to classify normal cases as well as acute, recent and old onset cases of MI, with AUROCs of 96.7\%, 82.9\%, 68.6\% and 73.8\%, respectively.  We have demonstrated a multi-lead fusion approach to detect the presence and occurrence-time of MI.  Our end-to-end framework provides flexibility for different levels of multi-lead ECG fusion and performs feature extraction via transfer learning.

\end{abstract}


}

\maketitle

% is not selected <OR> if conference mode is selected - because compsoc
% conference papers position the abstract like regular (non-compsoc)
\IEEEdisplaynontitleabstractindextext

% \fi
\IEEEpeerreviewmaketitle

\ifCLASSOPTIONcompsoc
\IEEEraisesectionheading{\section{Introduction}\label{sec:introduction}}
\else
\section{Introduction}
\label{sec:introduction}
\fi

\IEEEPARstart{C}{ardiovascular} diseases (CVDs) are the leading cause of death globally, and four out of five CVD deaths are due to heart attacks (i.e., myocardial Infarction) and strokes\cite{who2018}. Traditional diagnosis of heart attack mainly employ interpretation of ECG recordings, which requires precise acquisition devices and highly trained clinicians (i.e., cardiologists), both of which are in limited supply in resource-constrained areas. Cardiologists visually inspect the conventional 12-lead ECG waveforms as images when making diagnosis. However, such a process is tedious and can be highly subjective~\cite{zhu2014crowd}. ECG readings are also sensitive to mounting position and prone to movement artefacts~\cite{Martis2014}, resulting in noisy readings that add to the difficultly of making reliable diagnoses from them. {This problem becomes more visible in low-resource settings where there is no/limited cardiologists}. Numerous studies have also shown that it is not always possible to detect cardiovascular abnormalities from a visual inspection of the ECG trace alone, given the small amplitude and short PQRST durations involved \cite{Jahmunah2019}. In this context, computer aided diagnosis methods present a promising solution to the problem of analysing and identifying CVDs from ECG readings. 

Automatic approaches using computer algorithms have been proposed to extract domain-specific handcrafted ECG features both in time- and frequency-domains for heart disease diagnosis~\cite{variability1996standards,guidera1993signal,mehta2009detection, dash2009automatic}.  {(Compared to automatically learned features),Though handcrafted features,  such as  identifying complexes in the ECG trace, are simple and tractable, they might be susceptible to noise/motion artefacts and  do not generalise across variations in patient characteristics, mounting positions and device specifications}. Moreover, these approaches often require extensive pre-processing steps that are prone to error. Feature engineering also requires careful consideration and is associated with a time-consuming and labour-intensive model development process. By contrast, end-to-end deep learning methods can eliminate the need for explicit feature engineering, by learning optimal representations directly from the raw data for the task at hand. Additionally, there are currently no studies that have investigated the prediction of occurrence-time in MI (i.e., the age of a MI), information that is crucial for preoperative risk assessment~\cite{fleisher2014}: patients with acute (i.e., within seven days) and recent (i.e., less than 30 days but longer than seven days) MIs are considered to be at higher risk of a perioperative cardiac event, while those with old MIs (i.e., more than 30 days) are at higher risk of perioperative cardiac morbidity. {Such systems become handy, again, in low-resource settings, where there is no patient data archival practice; or in emergencies when a patient history might not be available on the spot.}

\begin{figure*}[t]
        \centering
			\includegraphics[scale=0.3]{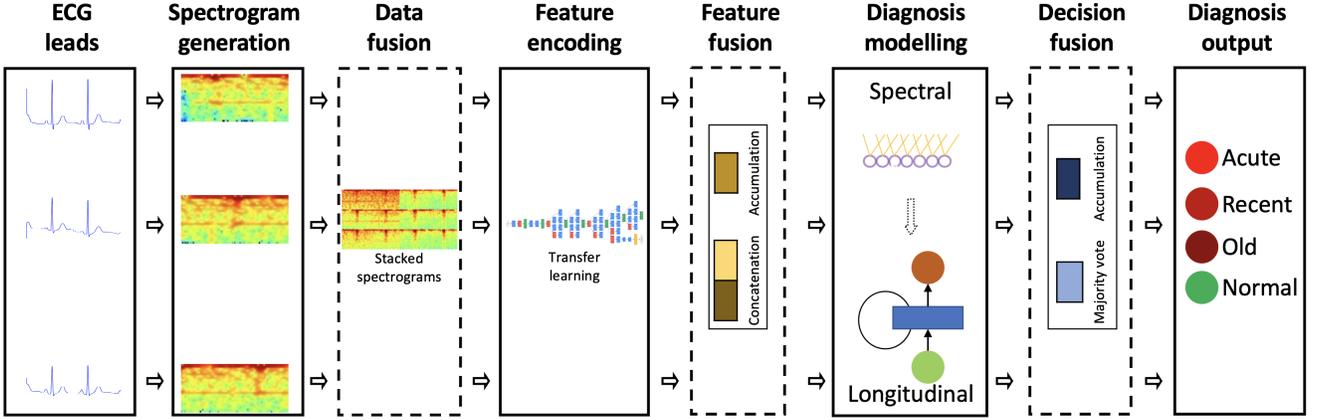}
		\caption{Overview of the proposed framework: multi-lead ECG waveforms are provided as input, and spectrogram generation is used to encode the frequency-time characteristics. Transfer learning is applied to encode deep features using existing computer vision networks. Spectral and temporal models were employed  for diagnosis modelling.}\label{fig:proposed_overview}
        \end{figure*}

In this paper, we present an end-to-end deep learning system for predicting the occurrence time of MI using 12-lead ECG waveforms (see Fig.~\ref{fig:proposed_overview}). 
The contributions of the proposed approach are as follows: 1)~spectral-based preprocessing that transforms time-series waveforms to spectrogram representations, which offer a more general representation that overcome issues of variability in sampling rates and device specifications across ECG manufactures; 2)~application of cross-domain transfer learning between natural images classification and ECG waveform detection, to extract spectral features to reduce the dimension of the spectrograms and minimise redundant information; 3)~ultilising joint spectral-temporal modelling to encode the spatial and temporal information from multi-channel ECG waveforms; 4)~proposing the use of a variety of fusion approaches to combine information at different levels of data representation (i.e., data, feature and decision level fusions) aiming to utilise distinctive characteristics offered by each ECG lead; and 5)~predicting the occurrence time of MI, utilising a large scale dataset containing $>15,000$ patients and $>323,000$ data samples, to provide timely intervention that can potentially improve patient outcomes. 

 The paper is organised as follows: Section~\ref{sec:related_work} surveys the research literature on the topic of automated ECG analysis for CVD diagnosis, with a focus on MI detection.
%, and summarises the distinctiveness of the proposed approach.
Section~\ref{sec:proposed_method} presents the proposed framework, including different multi-lead fusion strategies and diagnosis modelling techniques. Section~\ref{sec:experiments} details the experimental setup including descriptions of the dataset used for validation, the set up of network architectures and parameters in the proposed approach, in addition to methods considered for comparison. Section~\ref{subsec:results} presents results and discusses important findings, whilst concluding remarks are provided in Section~\ref{sec:conclusion}.

%\IEEEpubidadjcol

\section{Related work}\label{sec:related_work}

The potential diagnostic value provided by ECG signals, particularly for the diagnosis of CVDs such as MI, has long been recognised by researchers~\cite{Jahmunah2019}. To date, the majority of automated methods proposed for detecting MI from ECG traces, have focused on identifying abnormalities in the morphology of the signals. These approaches often entail error-prone pre-processing steps, such as identifying ECG complexes and handcrafting features to learn from (e.g. \cite{Adam2018}). The latter requires careful consideration and is associated with a time-consuming and laborious model development process. Deep learning methods by contrast, remove the need for feature engineering through their ability to automatically learn optimal representations directly from the raw data. 

Given the success of deep learning in a wide range of application domains, a growing number of studies have investigated the use of end-to-end deep networks for ECG based CVD detection \cite{AlRahhal2016, Acharya2017a, Reasat2017, Xiao2018a, Strodthoff2018, Raghunath2019, Goto2019, Darmawahyuni2019, Baloglu2019, Han2020}. The most widely employed and often best performing architectures have been CNNs, e.g. \cite{Acharya2017a, Strodthoff2018, Baloglu2019, Han2020}. A smaller number of these studies considered RNNs \cite{Darmawahyuni2019, Feng2019}, whilst \cite{Goto2019, Feng2019} demonstrated that a cascaded CNN-LSTM model architecture achieved superior performance to either individual model architectures on their own. This provides supporting evidence to the broad fact that capturing local and more long-term temporal characteristics of ECG signals through joint modelling, as we consider in this study, is well motivated.

Whilst deep learning methods have the potential to achieve state-of-the-art performance, this often entails a lengthy training and laborious model development process, requiring copious amounts of training data. This is particularly true when proposing sophisticated architectures from scratch, which contain many parameters that must be learnt. To overcome these challenges in data and labour requirements, transfer learning can be exploited. Transfer learning aims to leverage models that are pre-trained, namely models that have been carefully developed for a different task and/or on a different dataset. For the task of CVD detection using ECGs, the works of \cite{Xiao2018a, Xiao2018} showed the effectiveness of cross-domain transfer learning through their use of Google's Inception image recognition model \cite{szegedy2015going}. In these works, raw ECG signal traces were treated as images to be analysed by the network, thereby directly emulating the analysis cardiologists perform. Our work differs from these, and other similar studies, in that we first map 1D ECG signals to 2D spectrogram representations, before treating the latter as an image to be analysed by a pre-trained computer vision network through transfer learning. We consider the spectrogram representation as input, rather than the image of the ECG signal itself, as the former allows spectral time-varying information present in the signal to be better visualised. Thereby potentially enabling more clinically relevant information for diagnosis to be captured and learnt by an ML model. In previous work by the authors, we have shown the effectiveness of spectrogram image representations of physiological signals, such as PPG, for ML based disease diagnosis \cite{abebe2019cardio, abebe2020multi}. 

More recently, the following are examples of studies that have employed end-to-end deep learning approaches for ECG based MI detection specifically \cite{Strodthoff2018, Acharya2017a, Reasat2017,  Baloglu2019, Feng2019, Han2020}. For the majority of these works, MI detection was framed solely as a binary classification problem (non-MI or MI), whilst in others MI localisation was the focus through a multi-class problem formulation. The latter concerned identifying different MI types characterised by the location of the blocked artery (e.g. posterior lateral MI), as an example, 10-classes of MI location are predicted in the work of \cite{Baloglu2019}. The utility of discriminating between such MI types is clear, however prediction of MI by time-onset, as we propose in this work, has not yet been researched. Characterising MI diagnosis by estimated occurrence time has clear practical implications, as early detection of MI would enable proactive care management and intervention strategies to be enacted.

Published research in CVD detection from ECG signals, has primarily  restricted itself to using a single lead \cite{AlRahhal2016, Acharya2017a}. In the studies where multiple ECG leads were considered, individual models were developed and evaluated for each lead separately \cite{Baloglu2019}, that is without any data or result fusion, strategies we investigate extensively in this work. Whilst the use of 12 leads ECG may usually be limited outside clinical settings, and single lead analysis is advantageous in certain contexts (for example remote monitoring of patients), the work of \cite{Bradley2019} highlights the prevalence and seriousness of in-hospital MI. In these contexts, patients can be consistently and conveniently monitored using the standard and inexpensive 12 lead ECG setup, which clinicians use in their entirety to make diagnoses. Importantly, research has shown that different ECG traces contain different predictive importance for MI diagnosis \cite{Strodthoff2018}, highlighting the value of considering multiple lead fusion for CVD prediction in general \cite{Liu2017a, Strodthoff2018, Goto2019, Chen2019}. % Works that have considered fusion strategies for different ECG traces have achieved superior results compared to those that have not

Research methods proposed to date for automatic CVD detection, have primarily developed and evaluated their methods on the open-source PhysioBank Physikalisch-Technische Bundesanstalt (PTB) ECG dataset \cite{Goldberger2000}, or on proprietary datasets (e.g. \cite{Goto2019}). High performance achieved on the detection of MI or MI types in the PTB dataset, may indicate that to assess the generalisability of developed methods, validation on new datasets is vital. To date, studies that have employed alternative datasets have often been small. For example in \cite{Goto2019} models were trained on ECG records 10 seconds long from 362 patients with cardiovascular abnormalities. By contrast, we have developed our models using a dataset with over $17,000$ patients and $323,000$ ECG sample measurements, which is an order of magnitude higher than those used in the literature. Moreover, the dataset used in this work contain additional labels regarding the eventual time occurrence of MI, allowing us to investigate the possibility of predicting time-onset of heart attack from ECG readings alone.

%%%%%%%
\section{Proposed framework: DeepMI}\label{sec:proposed_method}

In  this  section,  we  outline  the  system  aspects  of  the  proposed  framework for identifying MI and its occurrence time. 

\subsection{Overview}
{Figure~\ref{fig:proposed_overview} shows the overview of our proposed approach. Given a dataset, $\mathcal{D}$, which contains $\eta$ patients diagnosed for MI, i.e., $\mathcal{D} =\{S_i\}_{i=1}^\eta$, where $\mathcal{S}_i$ represents the $i^{th}$ patient represented by $12$ ECG leads, $\mathcal{S}_i=\{\textbf{l}_i^j\}$, $j \in [I, II, III, aVR, aVL, aVF, V_1, V_2, V_3, V_4, V_5, V_6]$. Each patient is diagnosed and the onset time of MI cases is determined clinically. Thus, the ground truth of $\mathcal{S}_i$ is $g_i \in \{acute, recent, old, normal\}$. Thus,  our approach focuses on predicting the diagnose label using an end-to-end trainable framework designed to operate with limited training data and in low computational resource settings. To this end, we, first,   extract the frequency-time characteristics, i.e., spectrogram, of the ECG waveforms. Then feature encoding is done on these spectrograms via transfer learning using pre-trained computer vision networks. Diagnosis modelling is performed with spectral and longitudinal models, which encodes spatial and temporal information, respectively.  To integrate information from different ECG leads, we employ different fusion techniques. Next we describe the details of each step in the proposed framework.}

%%%
%\subsection{Preprocessing}\label{subsec:preprocessing}
Given the raw ECG waveforms, a preprocesing is applied  to clean up the data, which  includes filtering of patients with erratic ECG readings in one or more of the ECG leads. Particularly, missing readings which could occur due to loose mounting of the electrodes during ECG acquisition. {To discard some of the noise/motion artefacts, often characterized by low and high frequency characteristics, we employed    a band-pass filter (a high pass filter followed by a Gaussian filter) on all ECG recordings.} Other pre-processing steps include the subsequent sampling of ECG windows per lead, i.e., $\textbf{l}_i^j=[\textbf{w}_{in}^j]_{n=1}^\gamma$, where  $\textbf{w}_{in}^j$ represents the $n^{th}$ window segmented from the $j^{th}$ lead of the $i^{th}$ patient and  $\gamma$ is the total number of windows from a patient lead, $\textbf{l}_i^j$.

\subsection{Spectrogram generation}
Rather than using the raw ECG time-series to encode features for the MI diagnosis, the proposed framework employs a frequency-time (spectrogram) representation that captures the time-varying characteristics of the waveforms, and it is also robust across variations in device specifications such as sampling rate~\cite{ravi2016deep}. The spectrogram generation is computed from each $\textbf{w}_{in}^j$ by applying a Fourier transform, that results in a 2D frequency-time representation $F_{in}^j$. The subscript $i$ is dropped henceforth to improve readability. We also normalise spectrograms by their maximum value, followed by a logarithmic scale to smooth the representation as 
	\begin{eqnarray}
	\hat{F}_n^j=\log\left(\frac{F_n^j}{max(F_n^j)}*255\right).
	\end{eqnarray}
The smoothed $\hat{F}_n^j$ is then saved as image using the JPEG image format so that it resembles natural images for the transfer learning step. { While other colormaps could also be used, we employed the 'viridis' colormap to convert the spectrogram array to an image after normalization in Eq. (1). The JPEG format is selected due to its efficiency but other formats could also be used.}

        \begin{figure*}[h]
        \centering
             \includegraphics[width=0.2\textwidth]{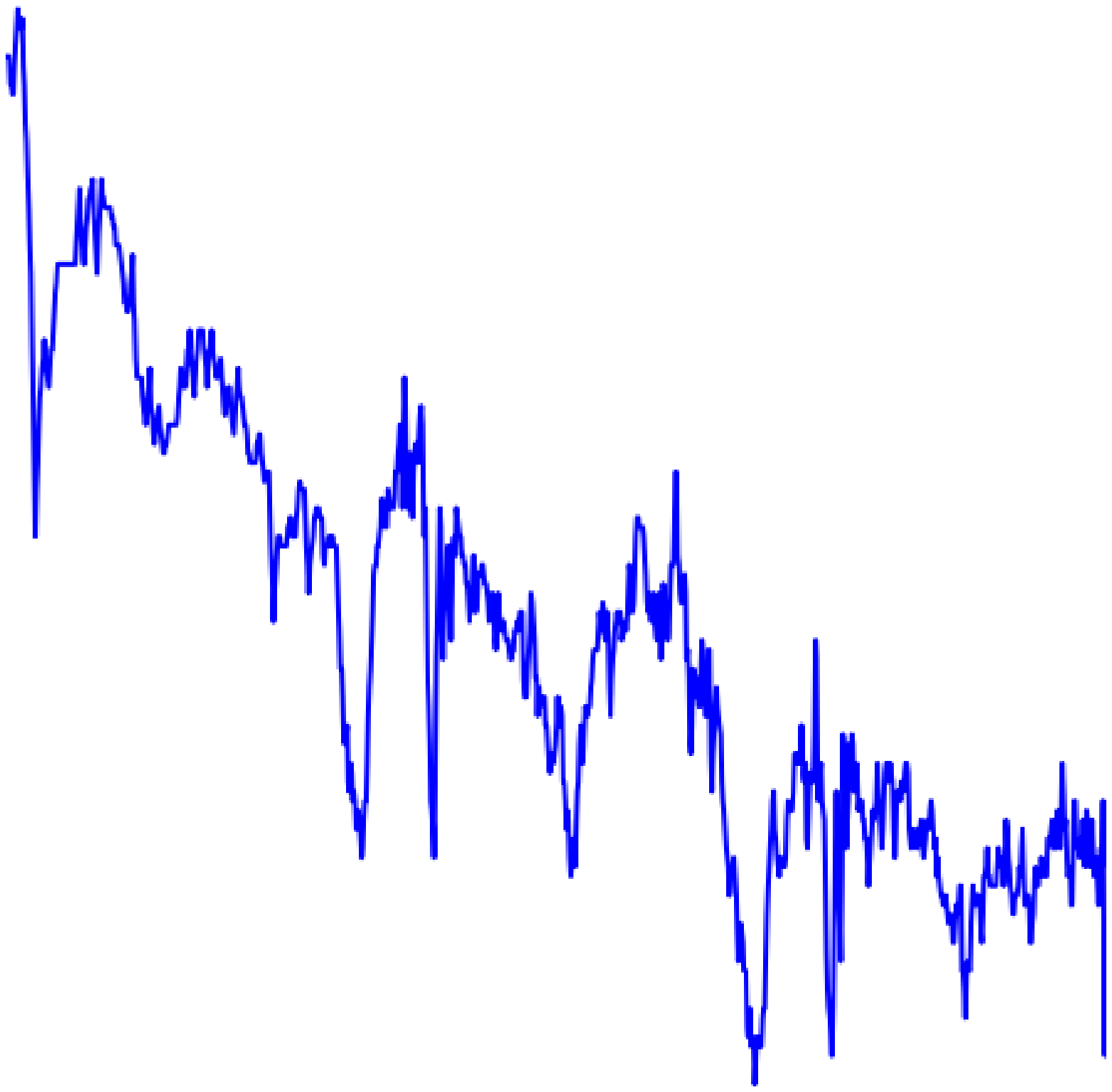}
             \includegraphics[width=0.2\textwidth]{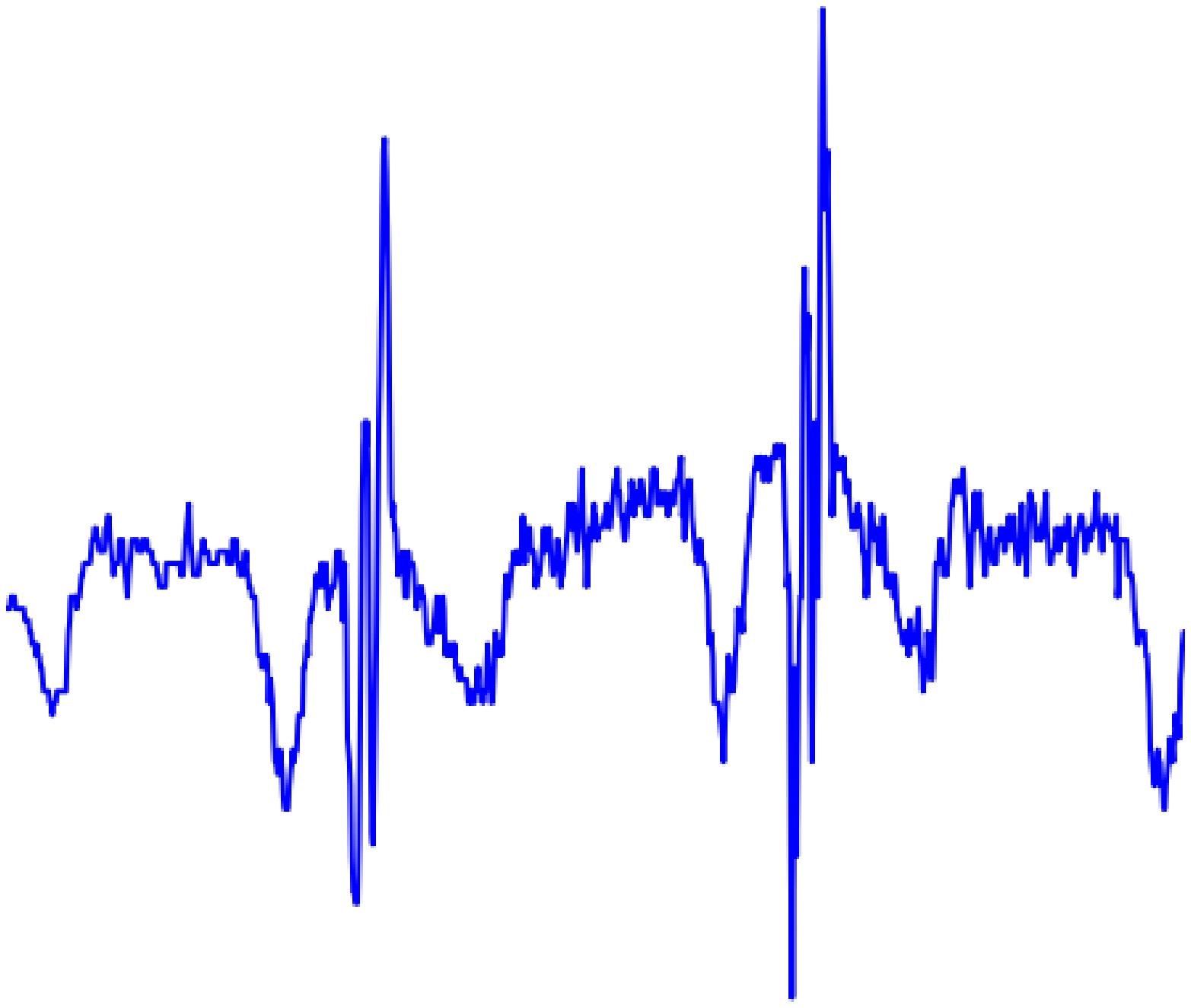}
             \includegraphics[width=0.2\textwidth]{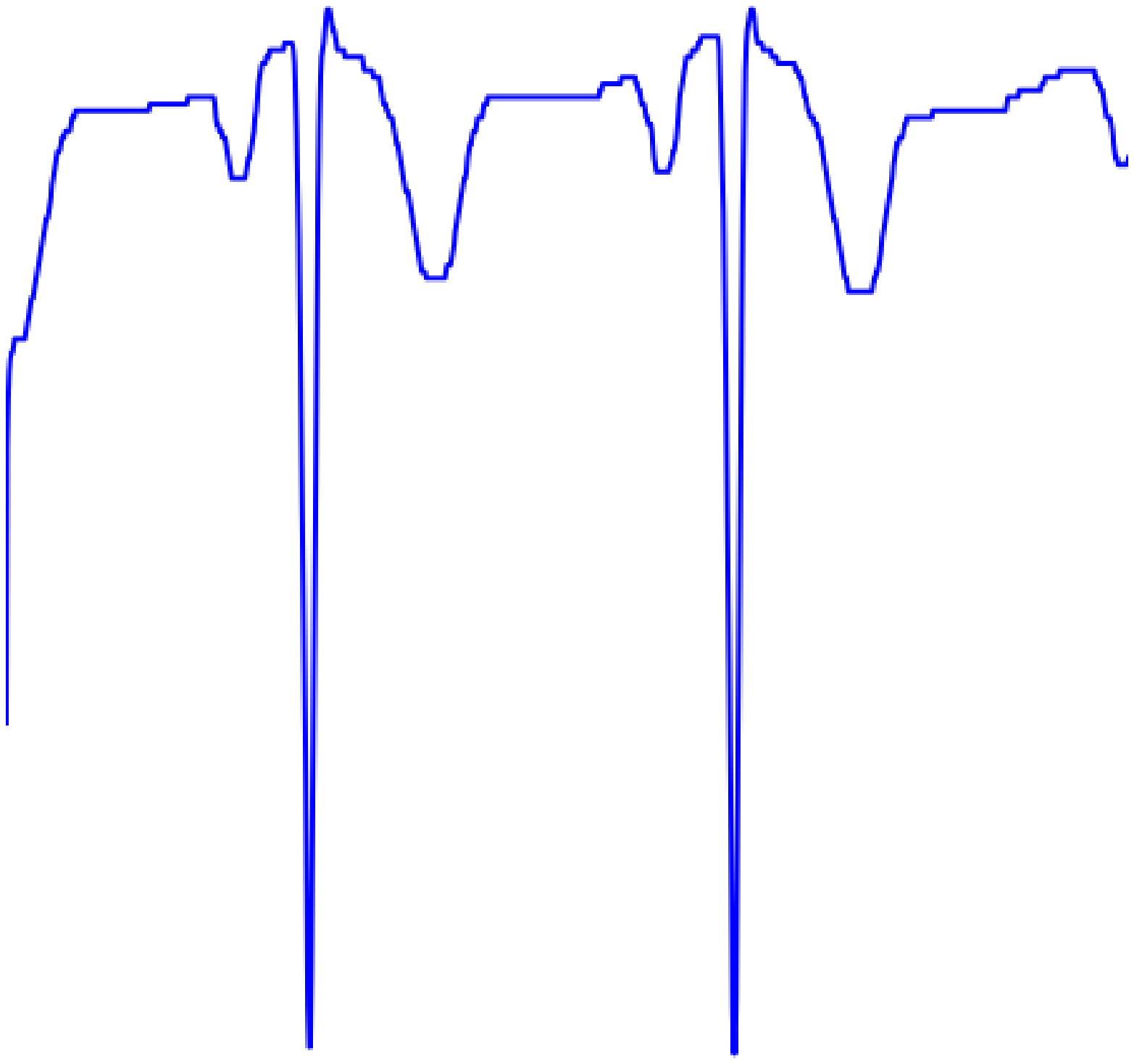}
             \includegraphics[width=0.2\textwidth]{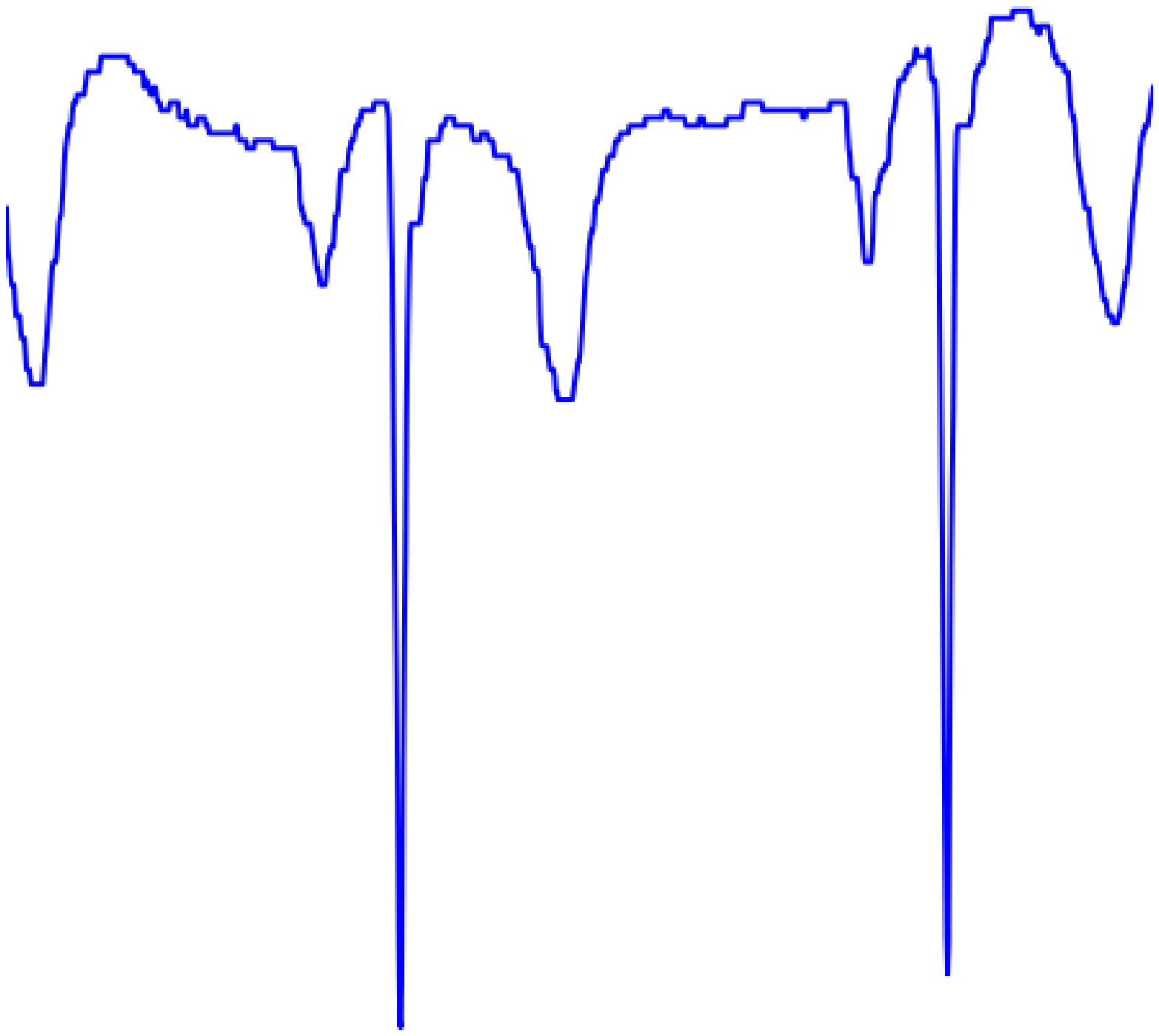}
			
        \subfloat[Acute]{
			\includegraphics[scale=1]{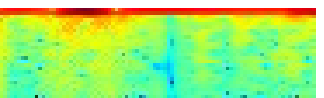}
		
			}\hspace{0.1cm}\hspace{0.1cm}
	        \subfloat[Recent]{
	       
			\includegraphics[scale=1]{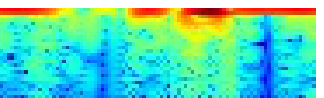}
			}\hspace{0.1cm}\hspace{0.1cm}
	        \subfloat[Old]{
	    \includegraphics[scale=1]{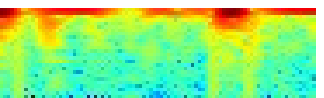}
			}\hspace{0.1cm}
	        \subfloat[Normal]{
	   
			\includegraphics[scale=1]{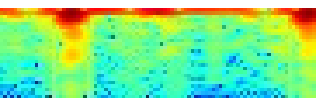}
			}
		\caption{Examples of ECG waveforms (top row) and their corresponding spectrograms (bottom row) from an aVR lead for a patient randomly selected from each class. Note: only 2-seconds duration of each waveform is shown to allow detailed visualisation of the patterns. }\label{fig:spec_examples}
        \end{figure*}

%%%%%%%%%%%%%%%%%%%%%%%%%%%%%%%%%%%%%%%%%%%%%
%%%

\subsection{Feature encoding}\label{subsec:hierarchical}
The spectrogram generation provides an image-like representation that suits convolution-based deep learning encoding. However, designing a dedicated deep network and encoding features by training it from scratch poses an enormous training data requirement, computational resources to train and a prolonged hyper-parameter tuning process. As a result, we opt to utilise existing computer vision networks, such as GoogLeNet~\cite{szegedy2015going} and MnasNet~\cite{tan2019mnasnet}, to encode features via a transfer learning approach. Thus, a hidden layer of these existing CNNs could be extracted from each normalised spectrogram input, $\hat{F}_n$, resulting a high-dimensional feature vector, $\textbf{d}_n \in \mathbb{R}^\tau$. The use of such cross-domain learning exploits powerful existing architectures and relaxes the training data requirement.

%%%%%%%%%%%%%%%%%%%%%%%%%%%%%%%%%%%%%%%%%%%%%
%%%
\subsection{Multi-lead fusion}\label{subsec:data_fusion}
The proposed framework offers a variety of fusion techniques to utilise the discriminant characteristics offered by the conventional 12 ECG leads~\cite{Strodthoff2018}. To this end, \textit{data}, \textit{feature} and \textit{decision} fusions are experimented with, details of which are provided below.

\subsubsection{Data fusion} The spectrograms from different ECG leads are fused together at  data level in the form of stacked spectrograms. Given the set of normalised spectrograms of the leads, $\{\hat{F}_n^j\}$ for $j \in[I, II, III, aVR, aVL, aVF, V_1, V_2, V_3, V_4, V_5, V_6]$, the data fusion is applied as in Equation~(\ref{eq:data_fusion}), which mimics the visualisation of the ECG leads by cardiologists in clinical practice~\cite{abebe2019cardio}. This results in a stacked spectrogram,  $\Phi_{n}$, as the output of the data fusion. 
\begin{eqnarray}
\Phi_{n}=\begin{bmatrix}
 \hat{F}_{n}^I,&\hat{F}_{n}^{aVR},& \hat{F}_{n}^{V1},&\hat{F}_{n}^{V4}\\
    \hat{F}_{n}^{II},&\hat{F}_{n}^{aVL},&\hat{F}_{n}^{V2},&\hat{F}_{n}^{V5}\\
    \hat{F}_{n}^{III},&\hat{F}_{n}^{aVF},&\hat{F}_{n}^{V3},&\hat{F}_{n}^{V6}
\end{bmatrix}\label{eq:data_fusion}.
\end{eqnarray}
In case of data fusion, transfer learning is applied to encode a single deep feature vector, $\textbf{d}_n$, from the stacked spectrogram, $\Phi_n$.

%%%
\subsubsection{Feature Fusion}\label{subsec:feature_fusion}
Another information fusion strategy from the multiple ECG leads could be implemented at the feature level. Namely, after deep features are extracted from the spectrogram of each lead, i.e.,  $\textbf{d}_n^j$ for $j  \in[I, II, III, aVR, aVL, aVF, V_1, V_2, V_3, V_4, V_5, V_6]$, where each $\textbf{d}_n^j$ is extracted from the corresponding $\hat{F}_n^j$. We have experimented with two feature fusion approaches: \textit{concatenation} and \textit{accumulation}. Concatenation refers to the stacking of the feature vectors to a single vector, i.e. the input to the modelling step becomes $\bar{\textbf{d}}_n=\mathcal{C}(\textbf{d}_n^j)$, where $\mathcal{C(\cdot)}$ represents the concatenation operation and, as a result, $\bar{\textbf{d}}_n \in \mathcal{R}^{N_l\times\tau}$, where $N_l$  represents the number of leads (i.e., $N_l = 12$)  and $\tau$ represents the dimension of each $\textbf{d}_n^j$. On the other hand, accumulation-based feature fusion, $\mathcal{A(\cdot)}$, results in $\hat{\textbf{d}}_n \in \mathcal{R}^{\tau}$, with equal dimension as the individual feature vectors due to the accumulation operation being defined as:
\begin{eqnarray}
\hat{\textbf{d}}_n  = \mathcal{A}(\textbf{d}_n^j) = \frac{\sum_{l=1}^{N_l}\textbf{d}_n^j}{N_l}.\label{eq:accumulation}
\end{eqnarray}

\subsubsection{Decision Fusion}\label{subsec:decision_fusion}

Decision fusion aims to exploit the distinctiveness in each lead by training a specific diagnosis model for each lead, after which the final output is obtained from the fusion of predictions from these lead-specific models. Given 12-lead ECGs, the diagnosis prediction specific to each lead results in $\textbf{p}_n^j \in \mathbb{R}^{N_d}$, where $N_d$ is the number of diagnosis classes. To obtain the final prediction output from these different lead models, we propose two different decision fusion techniques: \textit{accumulation} and \textit{majority vote}. Decision accumulation is applied similarly to feature fusion as in Equation~(\ref{eq:accumulation}) and the average of lead-specific prediction $\textbf{p}_n^j$ is computed. Majority vote, on the other hand, selects the most frequently predicted diagnosis class, across the different model leads, as the final predicted diagnosis.

%%%
%%%%%%%%%%%%%%%%%%%%%%%%%%%%%%%%%%%%%%%%%%%%%
%%%
\subsection{Diagnosis modelling}\label{subsec:modelling}
After deep features are obtained in the transfer learning module, three diagnosis modelling techniques are developed in the proposed framework: \textit{spectral} and \textit{longitudinal} and joint \textit{spectral-longitudinal}. These techniques are described more detail below. 

\subsubsection{Spectral modelling}
This modelling approach only uses the spectral information, encoded as deep features, for diagnosis modelling. To do so, a dense layer is applied that takes the deep feature vector as input; which means $\textbf{d}_n$ during data fusion, $\bar{\textbf{d}}_n$ during concatenated feature fusion, $\hat{\textbf{d}}_n$ during accumulated feature fusion and $\textbf{d}_n^j$ during decision fusion. The output of the dense layer is $\textbf{r}_n \in \mathbb{R}^\kappa$, where $\kappa$ is the dimension of the dense layer. The dense layer helps to refine the spectral features and reduce the feature dimension (since $\kappa < \tau$) as          \begin{eqnarray}
        \textbf{r}_n=\sigma({W}_{rd}\textbf{d}_n+\textbf{b}_r),
\end{eqnarray}\label{eq:dense}where $\sigma$ is a non-linear activation function, $W_{rd} \in \mathbb{R}^{\kappa \times \tau}$ is the weight matrix linking the deep feature vector and the dense layer, and $\textbf{b}$ is the bias vector. Note that during feature concatenation, $\tau$ becomes $N_l=12$ longer. The prediction vector for the MI diagnosis, $\textbf{p}_n$, is obtained by applying a softmax function on the dense layer output, $\textbf{r}_n$, as
         \begin{eqnarray}
        {\textbf{p}}_n=\frac{e^{W_{sr}\textbf{r}_n}}{e^{W_{sr}\textbf{r}_n}+1}\label{eq:sigmoid},
        \end{eqnarray}
where $W_{sr} \in \mathbb{R}^{N_c \times \kappa}$ is the weight matrix linking the dense layer and diagnosis classes and $N_c$ is the number of class labels. The diagnosis class corresponding to the index of the highest element in the prediction vector, $\textbf{p}_n$, becomes the final prediction output by the proposed framework.

%%%
\subsubsection{Longitudinal modelling}\label{subsec:sequential}
The spectral modelling above only encodes the spectral information in each sample window, $\textbf{w}_n$, segmented from a long ECG waveform. However, subsequent samples possess temporal dependency as $\gamma$ samples were segmented from each ECG lead. We propose a recurrent neural network to exploit these long-term temporal dynamics. Long short-term memory (LSTM) networks are designed to encode temporal dependency, and they handle the vanishing and exploding gradient problems better than the vanilla recurrent neural networks (RNNs) via their \textit{input}, \textit{output} and \textit{forget} gates that act as switches to control memory information about the past.

Given the  previous cell information, $\textbf{c}_{n-1}$, the output gate, $\textbf{o}_n$, the forget gate, $\textbf{f}_n$, the candidate cell information, $\bar{\textbf{c}}_n$, and input gate, $\textbf{i}_n$;  the current LSTM hidden state, $\textbf{h}_n$, can be computed as 
\begin{eqnarray}
\textbf{h}_n&=&\textbf{o}_n \odot \phi(\textbf{f}_n \odot \textbf{c}_{n-1} + \textbf{i}_n \odot \bar{\textbf{c}}_n)~\label{eq:lstm}, 
\end{eqnarray}
where $\odot$ is an element-wise multiplication and  $\phi$ is a $\tanh$ activation function. The dimension of  $\textbf{i}_n$, $\textbf{h}_n$, $\textbf{c}_{n-1}$, $\textbf{c}_n$,  $\textbf{o}_n$  and $\bar{\textbf{c}}_n$ is $\mathbb{R}^\nu$, which is the number of neurons in the LSTM layer. Note that when only the longitudinal model is used, its input is the high dimensional deep features. Finally, a softmax function similarly to Equation~(\ref{eq:sigmoid}) is applied on $\textbf{h}_n$ to obtain the MI diagnosis prediction, $\textbf{p}_n$, using the longitudinal model.

\subsubsection{Spectral-longitudinal modelling}
Separate use of the spectral and longitudinal modelling approaches have the following limitations. The spectral model fails to encode the long-term temporal dependency existing between subsequent samples segmented from  long duration ECG waveforms. On the other hand, the longitudinal model takes as input the high dimensional deep features ($\tau$ during data, feature accumulation and decision fusion, $12\tau$ during feature concatenation) obtained from the transfer learning step, which might result in overfitting due to the curse of dimensionality, especially when feature concatenation is used. We therefore propose a joint spectral-longitudinal model to address these limitations and utilise the advantages offered by both models, i.e. refinement of deep features and dimensionality reduction by the dense layer of the spectral model and the temporal encoding using LSTM in the longitudinal model. Thus, the output of the dense layer in the spectral model, $\textbf{r}_n$, is used as input to the longitudinal model. As a result, the hidden state, $\textbf{h}_n$, in the longitudinal model encodes the temporal dependency among subsequent $\textbf{r}_n$. Finally, the softmax layer is used to predict the MI diagnosis, $\textbf{p}_n$, similarly to Equation~(\ref{eq:sigmoid}).

%%%%%%%%%%%%%%%%
\begin{table*}[h]
    \centering
       \caption{The descriptions of methods employed and compared against each other in this study. {Different fusion techniques, such as \textit{Data}, \textit{Feature} and \textit{Decision} were experimented. Data fusion involves stacking of the spectrograms from different leads; feature fusion refers to the aggregation of features extracted from multiple leads. Decision fusion refers to the aggregation of decisions from different ECG leads. Two types of modelling were also employed: \textit{Spectral} and \textit{Longitudinal} that encode spatial and temporal information, respectively.}} \label{tab:comparison_method}
       	\resizebox{1\linewidth}{!}{
    \begin{tabular}{ll}
    \hline
     Method & Description \\ \hline 
     Data-Dense & Data fusion and  spectral modelling using dense layers\\
     Data-LSTM& Data fusion and  longitudinal modelling using LSTM\\
     Data-Dense-LSTM & Data fusion and  spectral modelling using dense layers and longitudinal modelling using LSTM \\ \hline \hline
    Accumulation-Feature-Dense & Accumulation-based feature fusion and  spectral modelling using dense layers\\
     Accumulation-Feature-LSTM & Accumulation-based feature fusion and  longitudinal modelling using LSTM\\
     Accumulation-Feature-Dense-LSTM & Accumulation-based feature fusion and  spectral modelling using dense layers and longitudinal modelling using LSTM \\ \hline
     Concatenation-Feature-Dense & Concatenation-based feature fusion and  spectral modelling using dense layers\\
     Concatenation-Feature-LSTM & Concatenation-based feature fusion and  longitudinal modelling using LSTM\\
     Concatenation-Feature-Dense-LSTM &Concatenation-based feature fusion and  spectral modelling using dense layers and longitudinal modelling using LSTM \\ \hline \hline
     Vote-Decision-Dense & Spectral modelling using dense layers and majority vote-based decision fusion\\
   Vote-Decision-LSTM & Longitudinal modelling using LSTM and majority vote-based decision fusion\\
    Vote-Decision-Dense-LSTM & Spectral modelling using dense layers, longitudinal modelling using LSTM, and majority vote-based decision fusion \\
   \hline
  Accumulation-Decision-Dense &Spectral modelling using dense layers and  accumulation-based decision fusion\\
     Accumulation-Decision-LSTM & Longitudinal modelling using LSTM and  accumulation-based decision fusion\\
     Accumulation-Decision-Dense-LSTM & Spectral modelling using dense layers, longitudinal modelling using LSTM, and  accumulation-based decision fusion \\ \hline
    \end{tabular}
    }
 
\end{table*}
%%%%%%%%

\section{Experimental Setup}\label{sec:experiments}
 In this section, we describe details of the dataset used for development and validation of onset time prediction, the set up associated with the different stages of the proposed framework. Namely, spectrogram generation, transfer learning, spectral and longitudinal model architectures and their corresponding hyper-parameters. We also describe the variety of methods used for comparative purposes.
%%%

\subsection{Dataset}\label{subsec:dataset}

	The ECGs from 17,381 patients (11,853 MI and 5,528 Normal cases) were collected in the Provincial Key Laboratory of Coronary Heart Disease, Guangdong Cardiovascular Institute (GCI), which is located in Guangdong province, China. The study has obtained ethics committee approval and informed patient consent. Each 12-lead ECG waveforms was anonymised, sampled at 500 Hz and was 10~s long. The ECG signals for each patient contain the standard 12 leads, which are \textit{I, II, III, V1, V2, V3, V4, V5,V6, aVF, aVL,} and \textit{aVR}. { Cardiologists annotated the MI cases furthermore into three sub-groups: \textit{Acute}, \textit{Recent} and \textit{Old} based on hospital records like patient history, in combination with ECGs.} The final GCI dataset resulted $1,489$ Acute (MI occurred within 7 days), $5,377$ Recent (MI occurred in less than 30 days but longer than 7 days) and $4,613$ Old (MI occurred beyond 30 days) MI cases.

%%%
\subsection{Parameter Setup}\label{subsec:parameter_setup}
% In this subsection, we describe the set-up of parameters used in computing the spectrogram, transfer learning, spatial and longitudinal modelling, and fusion parts of the proposed framework.

% \subsubsection{Preprocessing}
{We set the duration of a sample, $\textbf{w}_n$, to $1$ s in order to achieve a balance between longer duration (which might have some degree of redundancy and also reduce the number of training samples) and shorter duration (which might lack enough information to do inference).} We applied an overlapping  percentage of $50\%$, between subsequent samples, which results a total of $323,133$ samples from each lead, of which: $28,291$ are Acute, $102,163$ are Recent, $87,647$ are Old, and the remaining $105,032$ are Normal samples. In the spectrogram generation step, we applied a Fourier transform, with a chunk of $0.1$ s and an overlapping percentage of $90\%$, which results in a spectrogram representation with $91\times26$  resolution when stored as an image in JPEG format. 

% \subsubsection{Transfer Learning}
Among existing computer vision networks, we experimented with GoogLeNet's Inception-v3~\cite{szegedy2015going}  and MnasNet~\cite{tan2019mnasnet} networks that are trained on natural images from ImageNet~\cite{deng2009imagenet}. GoogLeNet is known to be quite robust due to its novel inception module and the fact that it has been validated across multiple domains. MnasNet, on the other hand, is considered a mobile computing friendly architecture, designed with limited resource settings in mind. Thus, we extract deep features from the penultimate layers of these networks using the spectrograms as input, resulting  $\tau = 2,048$ and $\tau = 1,056$ dimensional deep feature vectors from the Inception-V3 and MnasNet networks, respectively. 
%for each spectrogram. Different modalities are expected to contribute an equivalent number of samples from each patient. Thus, replication of samples is applied in order to achieve data balance among modalities when lower samples are available from a specific modality.
%\textbf{\subsubsection{Spectral and Longitudinal Modelling}}
After deep feature encoding using transfer learning, the spectral model employs a fully-connected dense layer with a dimension of $\kappa = 16$ neurons with a ReLu activation. We have aimed for an LSTM network which is simple and consists of a single layer of size $\nu = 16$. The number of time steps in the LSTM is set to $\gamma=19$ samples. Finally, the softmax layer consists a fully connected layer of size $N_c=4$, which is equal to the number of diagnosis classes. During training a \textit{sparse-softmax-cross-entropy} loss function is used with \textit{Adam} optimiser  and a learning rate of $0.01$.

{Both the GGH and PTB datasets used for the validation of the proposed framework exhibit imbalance of samples among classes. Though the number of MI cases is significantly higher than normal cases in GGH, the imbalance becomes smaller in the 4-class classification task OF MI onset detection since the MI cases further decomposed into Acute, Recent and Old classes. Considering the imbalance issue, we adopt the following strategies in our validation: a) we employ stratified cross-validation in  our training; b) every batch is designed to contain proportional number of samples from each class; c) sparse-softmax-cross-entropy is applied as our loss function that weighs the loss accordingly; d) Area under receiver operating characteristic (AUROC) is applied as the main metric to compute the classification performance, as this metric better handles the potential class imbalances. AUROC for each class is obtained using a one-vs-all strategy, and overall AUROC is computed from the average of AUROCs of all the four classes. We also employed accuracy, precision, sensitivity, specificity and F-score to evaluate MI detection performance.  The confusion matrices are also provided to obtain more insights on the potential misclassification errors among the classes.}

%%%
%\subsection{Methods of Comparison}\label{subsec:methods}
The proposed framework consists of a variety of fusion techniques (data, feature and decision) and modelling approaches  (spectral, longitudinal and spectral-longitudinal). These were experimented with for every possible combination of fusion and modelling approaches. A summary of these methods is provided in in Table~\ref{tab:comparison_method}.
%Apart from the changes these fusion approaches bring to the size of the feature and decision space, all other system parameters remain the same across these methods.

%%%
\section{Results and Discussion}\label{subsec:results}
{In this section, we present the results obtained for the classification of MI cases from normal cases as well as the prediction of the onset time of MI (i.e. acute, recent and old), which was generally treated as a four-class classification problem. We also performed the comparison of the proposed framework with existing features and classifiers.}

\subsection{MI Detection}
{We first evaluated the performance of our proposed approach to detect MI cases from Normal cases using MnasNet features on the GGH dataset. To this end, Spectral,  Longitudinal and Joint Spectral-Longitudinal models were evaluated on the stacked spectrogram (data-fusion) of multiple ECG leads. The results are shown in Table~\ref{table:results_mi}, where an AUROC value of $85.2\%$ is achieved using the Joint Spectral-Longitudinal model, higher than the independent performance of the Spectral and Longitudinal models. }

% 	====================================================
% Data Fusion - Dense
% Accuracy : 0.702765  Precision : 0.866894  Recall : 0.668175  Specificity : 0.777712  F-score : 0.754672 
% Class : MI   AUC: 0.806173 
% Average  AUC: 0.806173 
% ====================================================
% Data Fusion - LSTM
% Accuracy : 0.732200  Precision : 0.838875  Recall : 0.753285  Specificity : 0.686514  F-score : 0.793779 
% Class : MI   AUC: 0.798335 
% Average  AUC: 0.798335 
% ====================================================
% Data Fusion - Dense + LSTM
% Accuracy : 0.725359  Precision : 0.897626  Recall : 0.675661  Specificity : 0.833039  F-score : 0.770986 
% Class : MI   AUC: 0.851958 
% Average  AUC: 0.851958 

%%for DEEPMYO revision
    	\begin{table}[t]
		\centering
		\caption{{Performance of the proposed framework validated to detect MI cases from Normal Patients in the GGH dataset. Acc.: Accuracy, Pre.: Precision,  Sen: Sensitivity, Spe.: Specificity, F: $F_{1}$-score }}\label{table:results_mi}
		\resizebox{1\linewidth}{!}{%
					\begin{tabu}{l|ccccc|c}
					
									 \multicolumn{7}{c}{Performance metrics ($\%$)} \\ 
					\hline
				 Methods  & Acc. & Pre. & Sen. & Spe. & F &	AUROC \\ \hline
				 Spectral &  $70.3$ & $86.7$& $66.8$& $77.8$ & $75.5$ & $80.6$\\
				 Longitudinal & $73.2$ & $83.9$ & $\textbf{75.3}$ & $68.6$ & $79.4$ & $79.8$\\
			     Spectral-longitudinal & $72.5$ & $\textbf{89.8}$& $67.6$&   $\textbf{83.3}$& $77.1$& $\textbf{85.2}$\\ \hline

			\end{tabu}

		}
	\end{table}

 \subsection{Onset Time Detection of MI}
Table~\ref{tab:all_results} shows the AUROC (\%) values per each diagnosis class across different fusion approaches: data, feature and decision. Overall, feature fusion was shown to outperform the other fusion approaches as the majority of the classes achieved their highest performance through feature fusion approaches. Using GoogLeNet features (top half of Table~\ref{tab:all_results} the following highest AUROC values are reported for the classes: \textit{Acute} ($82.9\%$ using \textit{Concatenation-Feature-Dense-LSTM}), \textit{Normal} ($96.9\%$ using \textit{Concatenation-Feature-LSTM}) and \textit{Old} ($87.4\%$ using \textit{Concatenation-Feature-Dense-LSTM}). Using MnasNet features (bottom half of Table~\ref{tab:all_results}), all the four diagnosis classes achieved their highest AUROC values using feature fusion approaches, i.e., \textit{Acute} ($81.5\%$ using \textit{Concatenation-Feature-LSTM}),  \textit{Recent} (73.1\% using \textit{Concatenation-Feature-LSTM})), \textit{Normal} ($98.2\%$ using \textit{Concatenation-Feature-Dense-LSTM}) and \textit{Old} ($75.1\%$ using \textit{Concatenation-Feature-Dense-LSTM}). Results suggest that  \textit{Normal} cases are, as expected, easier to detect compared to MI cases. On the other hand, MI cases are shown to be challenging to classify, particularly \textit{Recent} cases as they are prone to being misclassified between the two extremes: \textit{Recent} and \textit{Old}. The results reported with MnasNet features in the bottom part of Table~\ref{tab:all_results} are particularly encouraging, as competitive performance is achieved with GoogLeNet features. The MnasNet framework is relatively simple compared to GoogLeNet, with the lower dimensionality of MnasNet features ($\tau=1,056$) proving to be an effective representation, particularly as input for the longitudinal models which otherwise were prone to overfitting using the $\tau = 2,048$-dimensional GoogleNet features. This is shown by the fact that the two MI cases achieved the best performance with the \textit{Concatenation-Feature-LSTM} method using MnasNet features. 

Generally, the superior performance obtained using feature fusion approaches reflects the unique discriminative characteristics of each ECG lead, which need to be exploited during modelling rather than merging early before modelling using data fusion or lately using decision fusion after modelling. Comparison between two feature fusion techniques: \textit{accumulation} and \textit{concatenation}, reveals both are effective but \textit{concatenation} is superior as the information corresponding to the distinctiveness of  each lead is fed into the model while \textit{accumulation} might lose this information. Although it is the simplest and least resource demanding, data fusion is shown to be the worst performing fusion technique, undoubtedly as the discriminative characteristics of the leads is lost too early before feature encoding. 

\begin{table}[t]
    \centering
       \caption{Performance of data, feature and decision fusion approaches in the proposed framework for the detection and classification of MI cases using spectral (Dense), longitudinal (LSTM) and spectral-longitudinal (Dense-LSTM) models, using: GoogLeNet (top table)and MnasNet features (bottom table). \textbf{A}: Acute,  \textbf{R}: Recent,   \textbf{N}: Normal and  \textbf{O}: Old}~\label{tab:all_results}
    \resizebox{1\linewidth}{!}{
    \begin{tabular}{lcccc|cc}\\
   % \vspace{-0.25cm} \\
     \multicolumn{7}{c}{\textbf{GoogLeNet features}} \\
      \hline
       &\multicolumn{4}{c}{\textbf{AUROC per class} ($\%$)} & \multicolumn{2}{c}{\textbf{
    Global} ($\%$)} \\
    \textbf{Method} & \textbf{A} & \textbf{R} & \textbf{N} & \textbf{O} & \textbf{AUROC} & {\textbf{Accuracy}}\\ \hline 
  Data-Dense & 70.0 & 62.2 & 83.8 & 63.5 &69.9 & {}69.9\\
     Data-LSTM & 75.7 & 66.1  & 92.7 & 69.2 &75.9 &  {73.4} \\
     Data-Dense-LSTM  & 78.1 & 53.9 & 85.9 & 67.9  &71.4 &  {70.3} \\  \hline
Accumulation-Feature-Dense  &  76.4 & 64.4 &86.4 & 64.1 &72.8 &  {71.3} \\
     Accumulation-Feature-LSTM & 73.1 & 69.5 & 92.1 & 69.2 &76.0 &  {73.4} \\
     Accumulation-Feature-Dense-LSTM & 77.8 & 65.7 & 93.9 & 72.2 &77.4 &  {75.6} \\ \hline
     Concatenation-Feature-Dense & 80.8 & 66.6 & 95.0 & \textbf{87.4} &77.8 &  {75.0}\\
    Concatenation-Feature-LSTM &62.0 & 64.1 & \textbf{96.9} & 68.8 &73.0 &  {68.7}\\
     Concatenation-Feature-Dense-LSTM & \textbf{82.9} &  68.6&  96.7  & 73.8 & \textbf{80.5} &  {\textbf{77.1}}\\ \hline 
    Accumulation-Decision-Dense & 74.0 &65.2 &85.3 &64.0 &72.1 &  {71.0}\\
     Accumulation-Decision-LSTM & 68.8 & \textbf{70.6} & 89.9 & 65.9 &73.8 &   {71.9} \\
     Accumulation-Decision-Dense-LSTM & 74. 8 & 67. 2 & 93. 2 & 71. 9& 76.8 &  {75.8}  \\ \hline \vspace{0.25cm} \\
    %  \multicolumn{6}{c}{} \\
     \multicolumn{7}{c}{\textbf{MnasNet features}} \\
     \hline
    &\multicolumn{4}{c}{\textbf{AUROC per class} ($\%$)} & \multicolumn{2}{c}{\textbf{
    Global} ($\%$)} \\
    \textbf{Method} & \textbf{A} & \textbf{R} & \textbf{N} & \textbf{O}  & \textbf{AUROC} & {\textbf{Accuracy}}\\ \hline 
     Data-Dense & 68.1 & 59.8 & 78.9 & 61.6 &67.1  & {68.9}\\
     Data-LSTM & 71.3 & 67.9 & 90.8 & 68.4  &74.6  & {73.3}\\
     Data-Dense-LSTM  & 73.7 & 54.8 & 77.6 & 59.8   &66.4  & {67.2}\\ \hline
     Accumulation-Feature-Dense  & 73.8 & 62.6 & 85.5 & 64.3  &71.5  & {70.9}\\
     Accumulation-Feature-LSTM & 72.1 & 68.0 & 91.0 & 68.5  &74.9   & {73.8}\\
     Accumulation-Feature-Dense-LSTM & 75.5 &  66.3 &  92.9 &  71.8 &76.6  & {73.7} \\ \hline
     Concatenation-Feature-Dense & 78.9 & 65.0 & 94.3 & 69.3 &76.9  & {74.6}\\
     Concatenation-Feature-LSTM &\textbf{81.5} & \textbf{73.1} & 97.7 & 74.7 &\textbf{81.8}  & {\textbf{78.3}}\\
     Concatenation-Feature-Dense-LSTM &  77.7 & 68.1 & \textbf{98.2} & \textbf{75.1}  &79.8  & {77.8} \\ \hline 
      Accumulation-Decision-Dense & 71.4 &  63.8 &  81.4 &  63.3 &70.0    & {69.4}\\
     Accumulation-Decision-LSTM & 70.3 & 68.6 & 90.8 & 67.1   &74.2  & {74.7}\\  
     Accumulation-Decision-Dense-LSTM & 73.1 &  66.9 &  91.9 &  70.2   &75.5  & {72.9}\\ \hline
    \end{tabular}
    }
 
\end{table}

% \subsection{Modelling approaches}
The overall results shown in Table~\ref{tab:all_results} provide further insights into the effectiveness of the proposed joint spectral-longitudinal modelling approach across the majority of the fusion techniques. The average AUROC performance show that joint spectral-longitudinal modelling achieved the highest in \textit{Feature Accumulation} ($77.4\%$), \textit{Feature Concatenation} ($80.5\%$), \textit{Decision Accumulation} ($76.8\%$) using GoogLeNet features. Similarly, superior performance by the spectral-longitudinal model is achieved using  \textit{Feature Accumulation} and \textit{Decision Accumulation} using MnasNet features. Separately, longitudinal models achieved the highest in data fusion technique ($75.9\%$ using GoogLeNet and $74.6\%$ using MnasNet features), where the feature dimension is not too long to lead to the curse of dimensionality and model overfitting. {We also provided the accuracy metric in the last column, and the results showed similar behaviour  as  the  AUROC, i.e., concatenation-based feature  fusion superior performance, compared to the other fusion techniques. During the validation using  MnasNet  features,  the  LSTM-based  longitudinal  models  tend  to  perform  competitively  with  Dense-LSTM models, due to the smaller dimensionality of MnasNet features ($1056$) compared to GoogLeNet ($2048$), and hence less overfitting.}

Among the decision fusion approaches, both majority vote and decision accumulation showed competitive performance. However, the former showed slight superiority over decision accumulation as shown in Fig.~\ref{fig:decision_confusion} (more in the Appendix), for the proposed joint spectral-longitudinal model. This is partly due to the loss of fine details resulting from the averaging performed in accumulation fusion techniques. Furthermore, the confusion matrices reveal misclassification errors observed in the classification of MI cases. It is clearly visible that the Normal class has been distinctively classified without significant misclassification with other MI cases. On the other hand, more misclassification errors occur among MI types, i.e. acute, recent and old. 
In summary, though data-fusion seems ineffective in leveraging the distinctiveness of the leads, care must still be taken when considering feature concatenation or decision fusion strategies. Namely, constraints imposed by training resource considerations due to the curse of dimensionality that might arise for feature concatenation, or the need of independent modelling for each lead in decision fusion.

%Discussion why do we care about the onset time of an infarction
Our DeepMI framework not only demonstrates its feasibility in identifying an infarction from ECG readings alone, that is free from any clinical input, but it also indicates the predicted onset time of an infarction. Previous studies have shown that the occurrence time of an MI affects both the reinfarction rate, as well as mortality in surgery \cite{livhits2011risk, garg2015preoperative}. In particular, the risk of developing an adverse cardiac event has been reported to decrease from 32.8\% to 18.7\% when surgery occurs within 30 days of MI, compared to surgeries occurring 31-60 days post-MI \cite{garg2015preoperative}. Utilising DeepMI could therefore provide the age of an MI from patient ECG readings only, without requiring previous patient histories or more involved laboratory test results. Our proposed framework could therefore be utilised to assist clinicians in remote settings, by providing additional information to assess patient risk.

%%%%%%%%%%%CONFUSION MATRICES - TWO COMPARE THE TWO DECISION FUSION APPROACHES %%%%%%%%%%%%%%%%%%%%

	\begin{figure}[t]
		\centering
		%\centering
		
			\subfloat[GoogLeNet features]{
			
				\includegraphics[width=0.25\textwidth]{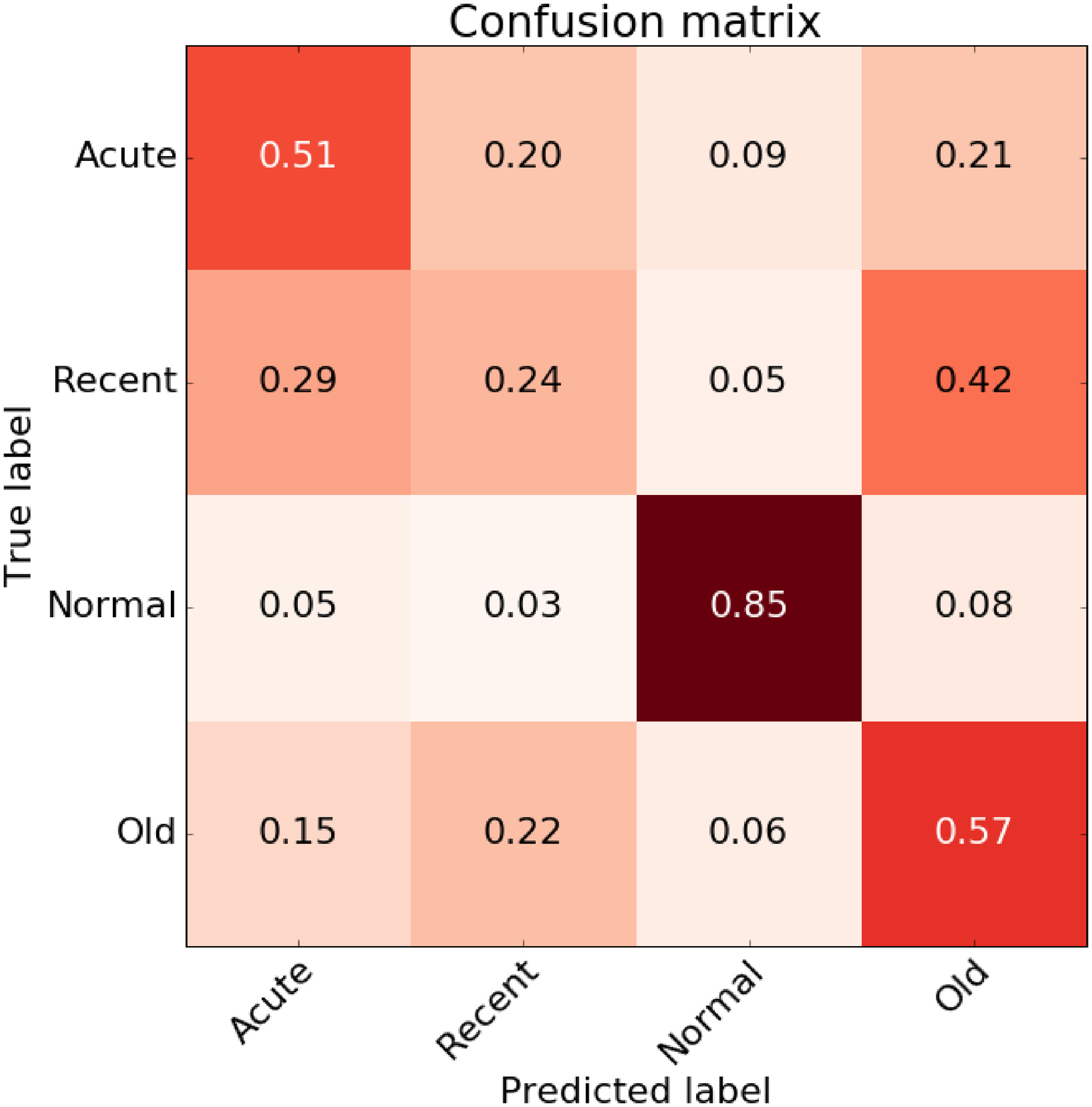}
			\includegraphics[width=0.25\textwidth]{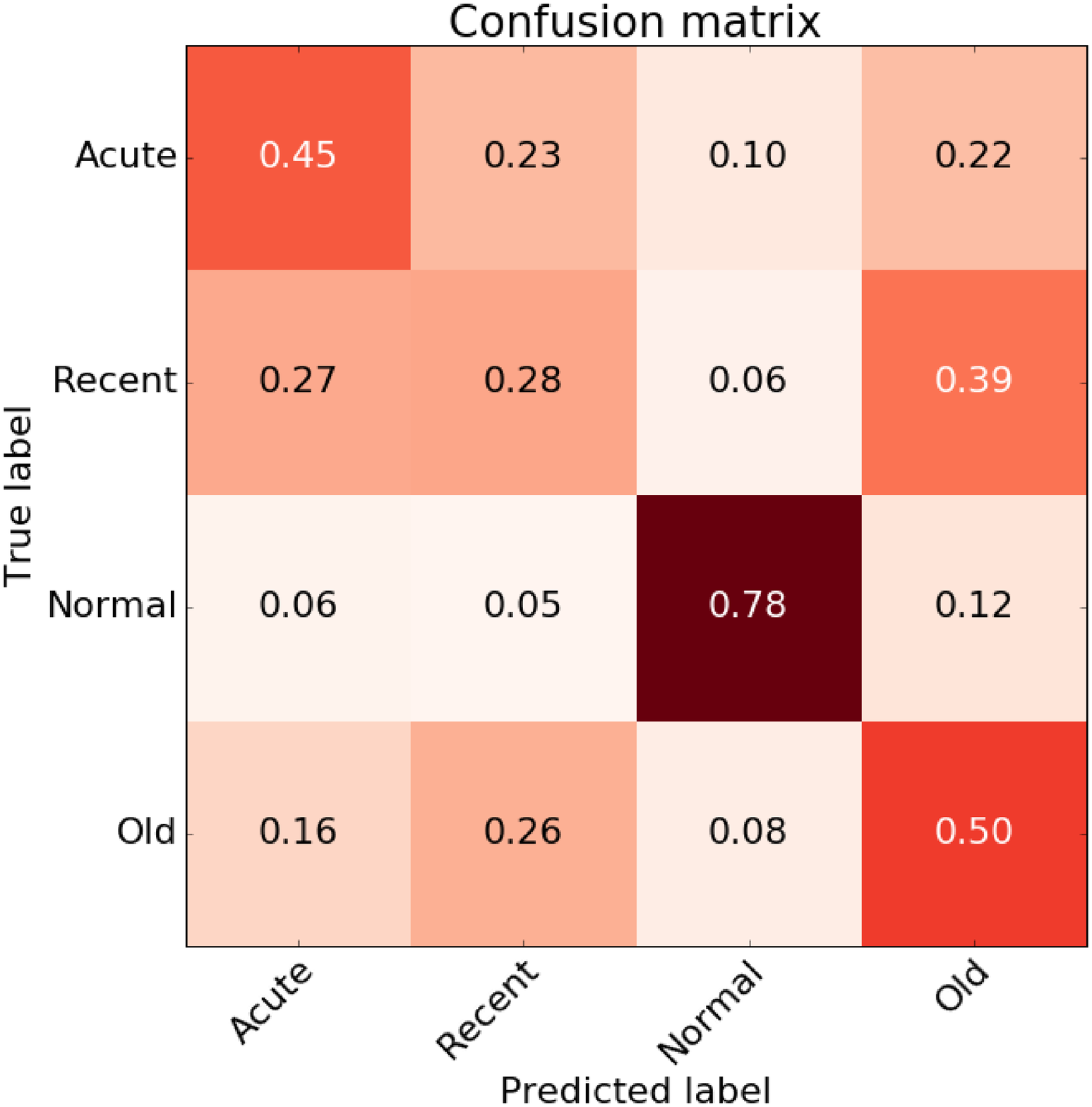}

			}
			
			\subfloat[MnasNet features]{
			\includegraphics[width=0.25\textwidth]{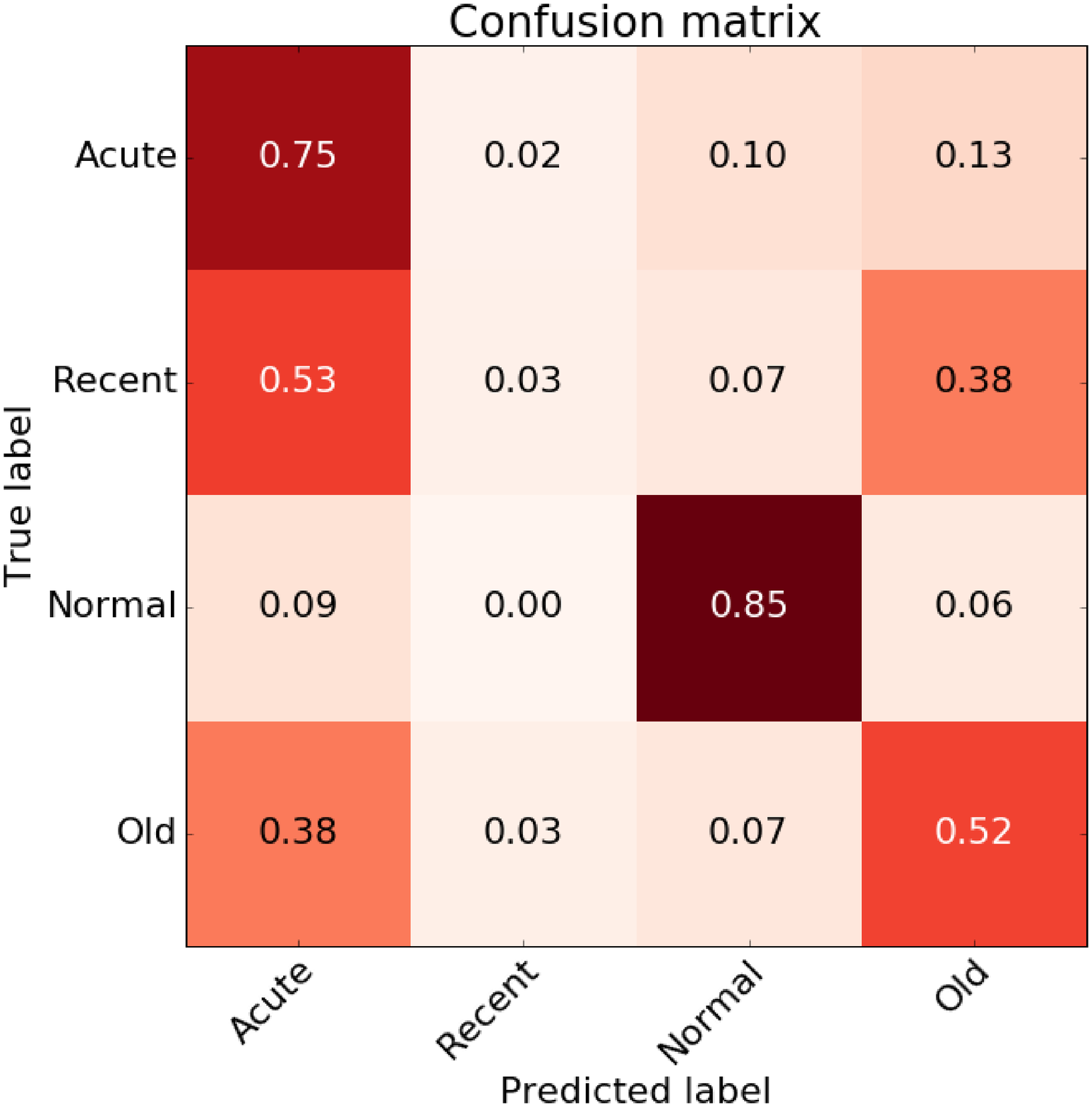}
						
				\includegraphics[width=0.25\textwidth]{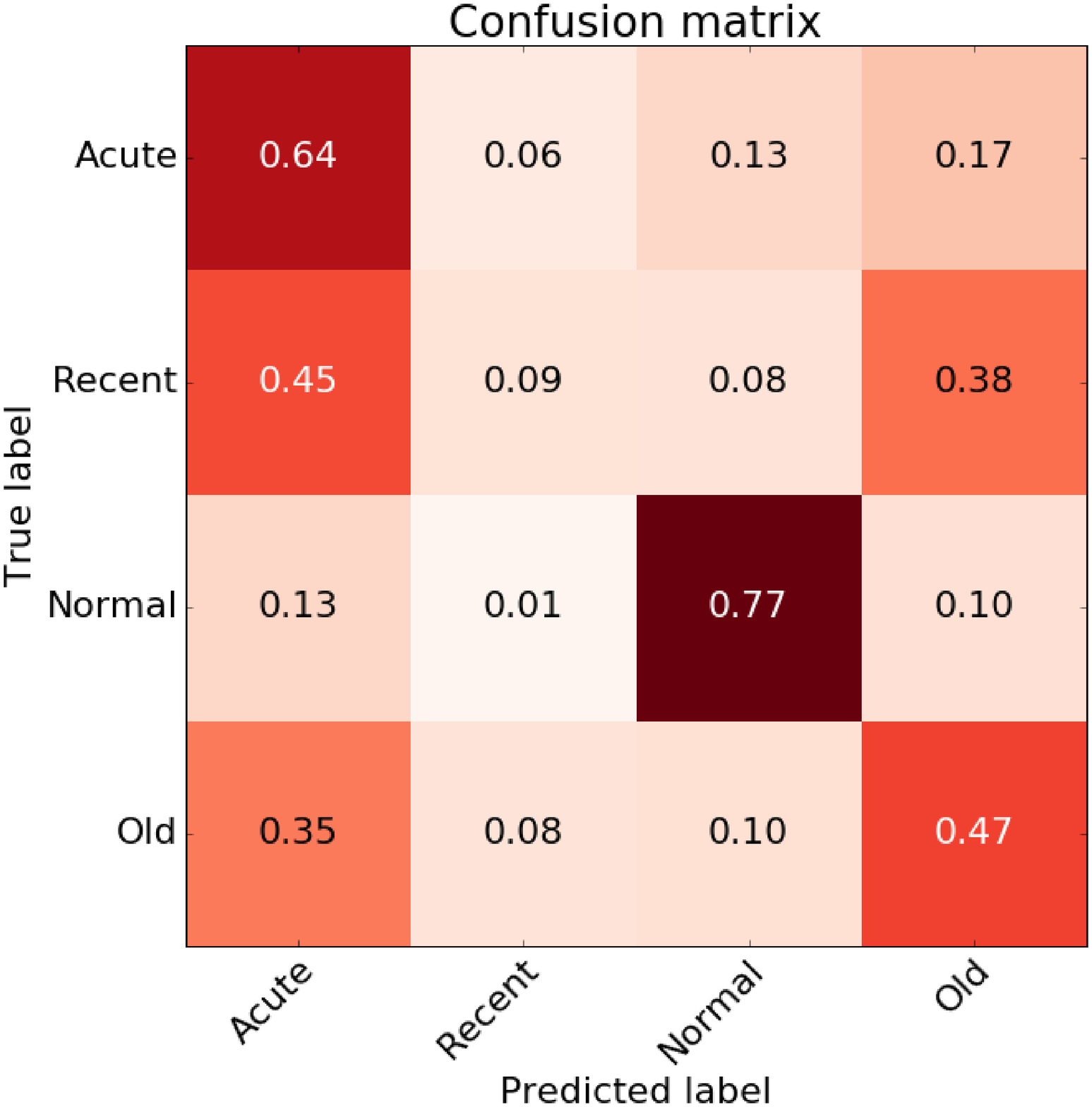}}
			\caption{Comparison of decision fusion approaches on (a) GoogLeNet and (b) MnasNet features. The first column contains \textit{Majority vote} results and the second column \textit{Decision accumulation}}\label{fig:decision_confusion}
		\end{figure}

\subsection{{Comparison with existing works and validation using a public database}}

 {
Among existing works, we selected hand-crafted features from ECG waveforms validated with common classifiers and dedicated deep neural networks~\cite{ranesparse2019}. The handcrafted features were obtained from autoregressive parameter estimation using Burg’s
method, which was developed for spectral estimation~\cite{bos2002autoregressive}, and it estimates the autoregressive coefficients optimised for the input signal, by minimizing loss in forward and backward prediction errors. Such representation was shown to outperform the fast-Fourier transform method in encoding dynamics of time-series signals, including ECG waveforms~\cite{al2014methods}. Common classifiers were used to validate on the autoregressive representation, which include Random Forest, Logistic Regression, K-nearest Neighbors, Gradient Boosted Trees and Multi-layer Perceptron. In addition to the handcrafted feature representation, simple deep networks, such as convolutional neural networks (CNNs) and Recurrent Neural Networks (RNNs) were implemented on the raw ECG waveforms. The CNN architecture adopted a four-layer architecture, and subsequent 1-D convolution was applied. The RNN architecture was made of a single-layer LSTM component, with a 1024 hidden units (using LSTM cells). Similar train-test validation was employed as of ~\cite{ranesparse2019}, i.e., 10-fold cross validation,  and the AUROC performance of both the existing works and the proposed (Dense, LSTM and Dense-LSTM) framework with the data-fusion approach were shown in Table~\ref{table:existing_work}. }

 {To further validate our proposed spectral-longitudinal model on a publicly available dataset, we used PhysioBank Physikalisch-Technische Bundesanstalt (PTB)~\cite{Goldberger2000}:  The PTB ECG dataset comprises 15-lead ECG records of  patients diagnosed with multiple heart diseases, sampled at 1000 Hz~\cite{goldberger2000physiobank,bousseljot1995nutzung}. In our evaluation, we only used 12-lead ECG data related to 148 (MI) and 52 (Healthy) subjects,  with a total of 200 subjects. As the duration of ECG data may vary across subjects, we used only the first 10-second segment of each patient.}

 {
The results showed that handcrafted features performed inferior to our approach while Gradient Boosted Tree achieved higher performance compared to other non-deep learning based classifiers. The deep networks, i.e., CNNs and RNNs, that were trained from scratch with the  raw ECG waveforms shown to underperform in their MI detection power, partly due to the limited amount of data available for training. This validates our approach which aims to utilize existing networks via transfer learning rather than designing a new network for each problem and train it extensively with a given dataset.
}%coloring

     	\begin{table}[t]
		\centering
		\caption{{AUROC (\%), results from the publicly available PTB database~\cite{Goldberger2000}, using existing works that employed hand-crafted feature, common deep learning classifiers, dedicated deep networks (CNN and RNN~\cite{ranesparse2019}). A comparison is conducted with data-fusion strategy of our proposed framework to detect MI cases from Normal cases.}}\label{table:existing_work}
		\resizebox{1\linewidth}{!}{%
					\begin{tabu}{lllc}

			    \multicolumn{4}{c}{} \\ 
						 \multicolumn{4}{c}{PTB Dataset} \\ 
					\hline
				  	 & Method & Classifier&	AUROC \\
				\hline
					 \multirow{7}{*}{Existing}&\multirow{5}{*}{Handcrafted} & Random Forest & $68\%$ \\
					 & & Gradient Boosted Tree & $70\%$ \\
					 & & Multi-layer Perceptron & $61\%$ \\
					 & & Logistic Regression & $66\%$ \\
					 & & K-nearest Neighbor & $64\%$  \\ \cline{2-4}
					 & \multirow{2}{*}{Deep Learning}& CNN & $49\%$ \\
					 & & LSTM & $49\%$ \\ \hline
					 \multirow{3}{*}{Proposed} & Spectral & Dense & $88\%$ \\
					 & Longitudinal & LSTM & $90\%$ \\
					 & \textbf{Spectral-Longitudinal} & \textbf{Dense-LSTM} &  $\textbf{94\%}$  \\ \hline
% 				\multirow{3}{*}{Data Fusion}  \\ \HLIN
% 				Spectral & $\textbf{99}\%$& $60\%$& $89\%$ & $88\%$\\
% 				Longitudinal &$96\%$& $\textbf{83}\%$& $85\%$ & $90\%$\\
% 			Spectral-longitudinal &$98\%$& $66\%$&  $\textbf{95}\%$& $\textbf{94}\%$\\ \hline

			\end{tabu}

		}
	\end{table}

\section{Conclusion}\label{sec:conclusion}
Myocardial infarction (MI) is globally the leading of cause of death among cardiovascular diseases. Due to the delay and resource expenditure associated with diagnosis from laboratory based blood sample tests, it is common clinical practice to inspect the electrocardiogram (ECG) records of patients instead. However, such an approach is still time consuming and subject to interpretation bias. In most developing countries, the number of cardiologists to interpret the ECG records is far lower than demand requires. In this context, data-driven approaches can assist the diagnosis process by providing decision support to the domain experts on site. Particularly, understanding the onset time of MI is crucially important in providing effective and early intervention, thereby improving patient outcomes. To this end, we have proposed an end-to-end trainable deep learning framework that takes raw ECG records and detects MI cases from normal (or non-MI cases), whilst also inferring the occurrence time of the heart attack as being either acute (within 7 days), recent (less than 30 days but longer than 7 days) and old (beyond 30 days). To do so, we employ a transfer learning technique that aims to exploit existing convolutional neural networks by representing the raw ECG time series with an image-like spectrogram representation. Importantly, we have experimented with a variety of fusion techniques to utilise the unique characteristics present among different ECG leads. We have proposed a joint spectral-longitudinal model, where the spectral component refines the high-dimensional deep features and applies dimension reduction to avoid the curse of dimensionality, and the longitudinal model encodes the temporal dependency among subsequent ECG samples. Encouraging performance, as high as $81.8\%$ AUROC, is obtained when we validated the proposed framework on a cohort of $>17,000$ Chinese patients. Though the onset-time detection performance could sometimes be less optimal, e.g., for \textit{Recent} MI cases, the proposed platform could still provide useful insights, such as perioperative risks, particularly in low-resource settings where there is very minimal cardiologists-to-patients  and/or rich patient history is not common/available.   Future work aims to expand the validation of the proposed framework to multiple recent networks, such as ResNeSt~\cite{zhang2020resnest} and to utilize attention-based mechanisms in the longitudinal modelling stage.

	\bibliographystyle{IEEEtran}
	\bibliography{DeepMI}

\end{document}